\documentclass[apj]{emulateapj}

\begin{document}

\newcommand\rxs{1RXS J154439.4$-$112820}
\newcommand\pos{(J2000) R.A.=$15^{\rm h}44^{\rm m}39^{\rm s}\!.38$, decl.=$-11^{\circ}28^{\prime}04^{\prime \prime}\!.3$}
\newcommand\psr{PSR J1023+1038}

\submitted{Accepted for publication in the Astrophysical Journal Letters on April 1, 2015}
 
\shorttitle{3FGL J1544.6$-$1125: A Transitional MSP}
\shortauthors{Bogdanov \& Halpern}

\title{Identification of the High-Energy Gamma-Ray Source 3FGL J1544.6$-$1125 as a Transitional Millisecond Pulsar Binary in an Accreting State}

\author{Slavko Bogdanov and Jules P.~Halpern}

\affil{Columbia Astrophysics Laboratory, Columbia University, 550 West 120th Street, New York, NY 10027, USA}

\begin{abstract}  
We present X-ray, ultraviolet, and optical observations of 1RXS
J154439.4$-$112820, the most probable counterpart of the unassociated
\textit{Fermi} LAT source 3FGL\,J1544.6$-$1125. The optical data
reveal rapid variability, which is a feature of accreting systems. The
X-rays exhibit large-amplitude variations in the form of fast
switching (within $\sim$10 s) between two distinct flux levels that
differ by a factor of $\approx$10. The detailed optical and X-ray
behavior is virtually identical to that seen in the
accretion-disk-dominated states of the transitional millisecond pulsar
binaries PSR\,J1023+0038 and XSS J12270$-$4859, which are also
associated with $\gamma$-ray sources.  Based on the available
observational evidence, we conclude that 1RXS J154439.4$-$112820 and
3FGL J1544.6$-$1125 are the same object, with the X-rays arising from
intermittent low-luminosity accretion onto a millisecond pulsar and
the $\gamma$-rays originating from an accretion-driven outflow. 1RXS
J154439.4$-$112820 is only the fourth $\gamma$-ray emitting low-mass
X-ray binary system to be identified and is likely to sporadically
undergo transformations to a non-accreting rotation-powered pulsar
system.
\end{abstract}


\section{INTRODUCTION}
Within the past two years, three neutron star binaries, PSR
J1824$-$2452I in M28 \citep{Pap13}, PSR J1023+0038 \citep{Pat14}, and
XSS J12270$-$4859 \citep{Bassa14}, have been observed to
switch between accreting and rotation-powered pulsar states. These
discoveries have confirmed the standard model for the formation of
rotation-powered millisecond pulsars (MSPs) via spin-up by accretion
in a low-mass X-ray binary (LMXB).

The two confirmed transitional MSPs in the field of the Galaxy, PSR
J1023+0038 and XSS J12270$-$4859, exhibit strong $\gamma$-ray emission
in \textit{Fermi} Large Area Telescope data even during their
accreting states. This implies that other similar objects may be
hiding among the unidentified \textit{Fermi} LAT sources \citep[see,
  e.g.,][for the case of 1FGL J1417.7$-$4407]{Stra15}.  While targeted
radio pulsation searches of \textit{Fermi} LAT sources have proven to
be particularly fruitful for discovering new rotation-powered MSPs,
yielding at least 38 published discoveries to date\footnote{See
  https://confluence.slac.stanford.edu/display/GLAMCOG/
  \\ Public+List+of+LAT-Detected+Gamma-Ray+Pulsars for an up-to-date
  list.}, transitional objects may elude detection at radio
frequencies since the pulsar emission mechanism appears to be
extinguished when an accretion disk is present in the system
\citep[e.g.,][]{Stap14}.

In their low-luminosity ($\sim$$10^{33}$\,erg s$^{-1}$ in the soft
X-ray band) accreting states, PSR J1023+0038, XSS
J12270$-$4859, and PSR J1824--2452I exhibit sudden, unpredictable
drops in flux and occasional intense flares
\citep{deM10,deM13,Lin14,Ten14,Bog14b}. This characteristic X-ray
behavior is a unique signature that can be used to identify additional
transition systems and distinguish them from other varieties of
accreting objects such as cataclysmic variables (CVs).

We have identified a particularly promising candidate for an LMXB/MSP
transition object, namely, 3FGL J1544.6$-$1125, previously known as
2FGL J1544.5$-$1126. The only X-ray-bright object within the 95\%
confidence error ellipse of this 3FGL source is the \textit{ROSAT}
source 1RXS J154439.4$-$112820 \citep{Step10}.  Optical identification of
\textit{ROSAT}-selected candidates for unassociated \textit{Fermi} LAT
sources by \citet{Mas13} revealed this object to have an optical
spectrum in 2012 consistent with those of accretion-disk-dominated
systems, with prominent emission lines of H and He. On this basis, it
was classified as a CV and was dismissed as the counterpart of 3FGL
J1544.6$-$1125 because no CVs are known to exhibit persistent
$\gamma$-ray emission.  However, given the fact that both PSR
J1023+0038 and XSS J12270--4859 were mis-classified as CVs upon their
discovery suggests that similar objects may also be masquerading as
CVs.  Based on this, we suspected that 1RXS J154439.4$-$112820 is
indeed the counterpart of the \textit{Fermi} LAT source. Here, we
present an analysis of X-ray and UV data from \textit{XMM-Newton} and
optical time-series photometry obtained at MDM Observatory of 1RXS
J154439.4$-$112820, which reveal the true nature of this object.

\begin{figure}
\begin{center}
\includegraphics[width=0.42\textwidth]{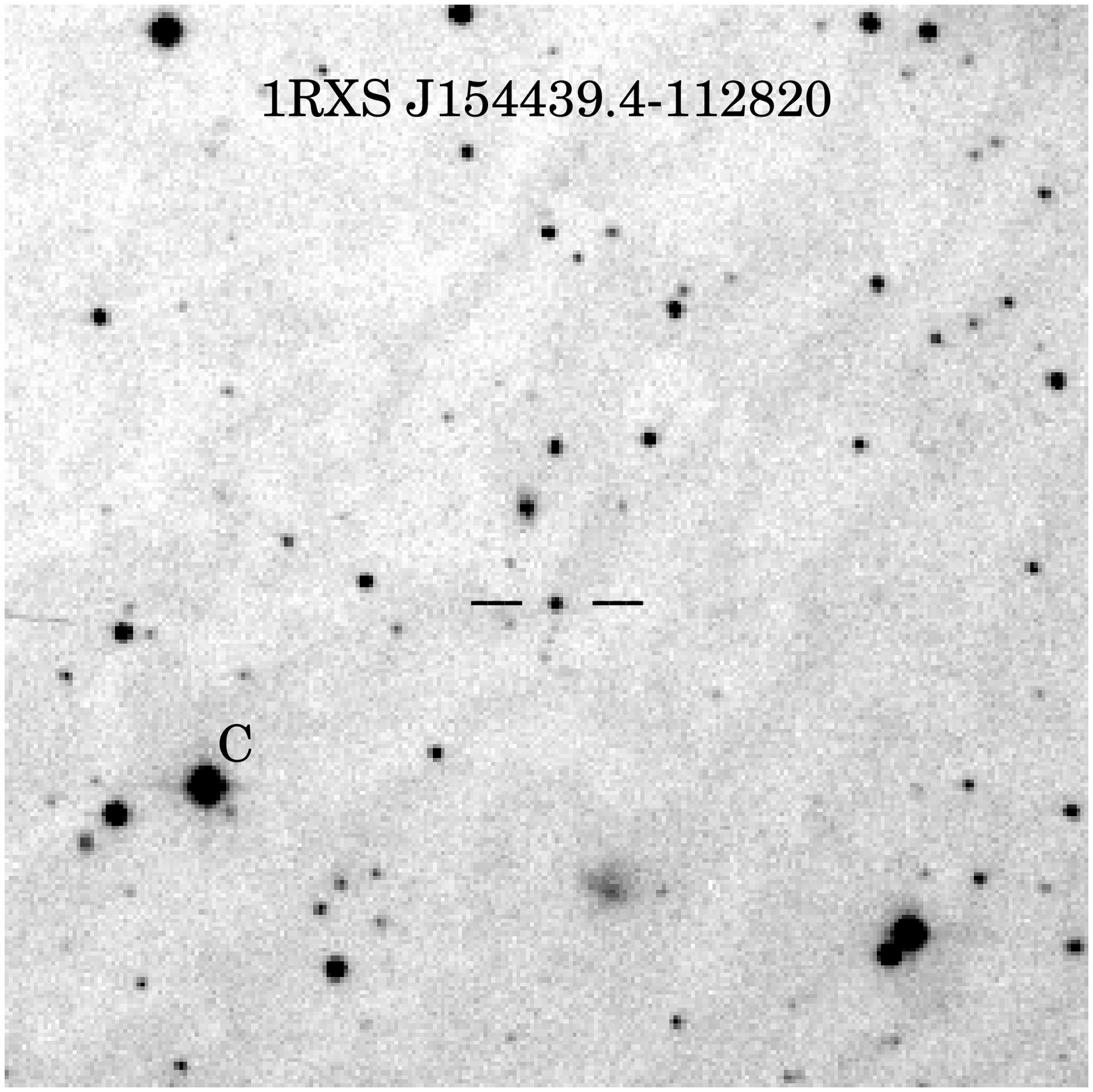}
\end{center}
\caption{Finding chart for \rxs\ (tic marks) at position \pos\ from
  MDM 1.3m images in the GG420 filter.  The size is
  $4^{\prime}\!.3\times4^{\prime}\!.3$.  North is up and east is to
  the left.  The star marked ``C'' has magnitude $R=14.25$ in the USNO
  B1.0 catalog, and is used for appoximate calibration in GG420.
  Faint fringing is a detector artifact due to the filter transmission
  at long wavelengths.  }
\label{fig:finder}
\end{figure}

\begin{deluxetable*}{cccccl}
\tabletypesize{\small}
\tablewidth{0pt}
\tablecaption{Log of MDM Observatory Time Series of \rxs\tablenotemark{a}}
\tablehead{
\colhead{Telescope} & \colhead{Date (UT)} & \colhead{Time (UT)} &
\colhead{Exp. time (s)} &  \colhead{$N_{\rm exp}$} & Conditions
}
\startdata
  2.4-m  &  2014 Mar 22  & 09:53--12:42  &  10  &  677  & Clear \\
  2.4-m  &  2014 Mar 23  & 07:15--12:40  &  10  & 1331  & Photometric \\
  1.3-m  &  2015 Feb 13  & 09:44--13:14  &  30  &  379  & Cirrus, near moon \\
  1.3-m  &  2015 Feb 14  & 09:45--13:13  &  30  &  376  & Thin clouds \\
  1.3-m  &  2015 Feb 17  & 09:19--13:13  &  30  &  425  & Partly cloudy \\
  1.3-m  &  2015 Feb 18  & 09:14--11:52  &  30  &  288  & Photometric 
\enddata
\label{tab:log}
\tablenotetext{a}{All observations in the GG420 filter.}
\end{deluxetable*}

\section{OBSERVATIONS AND DATA REDUCTION}

\subsection{XMM-Newton}
The field around 1RXS J154439.4$-$112820 was observed with
\textit{XMM-Newton} on 2014 February 16 (ObsID 072080101). The EPIC MOS1
instrument was configured in ``full frame'' mode The MOS2 and PN
detectors were used in ``timing'' mode, which provide time resolution
of 1.75 ms and 30 $\mu$s, respectively. In this mode, the fine time
resolution is enabled by imaging in only one spatial dimension (RAWX),
with the other dimension (RAWY/TIME) used for fast read out.  The
medium optical blocking filter was used for all three instruments.
The effective exposures for the MOS1, MOS2, and pn instruments were
40.9 ks, 40.6 ks, and 38.9 ks, respectively. All three data sets were
processed with the SAS\footnote{The XMM-Newton SAS is developed and
  maintained by the Science Operations Centre at the European Space
  Astronomy Centre and the Survey Science Centre at the University of
  Leicester.} version {\tt xmmsas\_20141104\_1833}.  The MOS1 source
events were extracted from a circular region of radius 36$''$. The
MOS2 and pn data were extracted from regions of width 60 and 7
detector pixels, respectively, in the imaging (RAWX) direction and the
full detector width in the readout (RAWY/TIME) direction. The standard
flag and pattern event screening criteria were applied to all three
data sets.

The \textit{XMM-Newton} Optical Monitor (OM) was used with the $U$
filter (centered on 3440 {\AA} with a band pass of 840 {\AA}) in place
and in "fast" mode, which permits rapid readout and photon counting
capabilities.  A photometric light curve was obtained using the {\tt
  omfchain} processing pipeline with the default parameter settings.

%
\begin{figure*}[!t]
\begin{center}
\includegraphics[height=5.8cm]{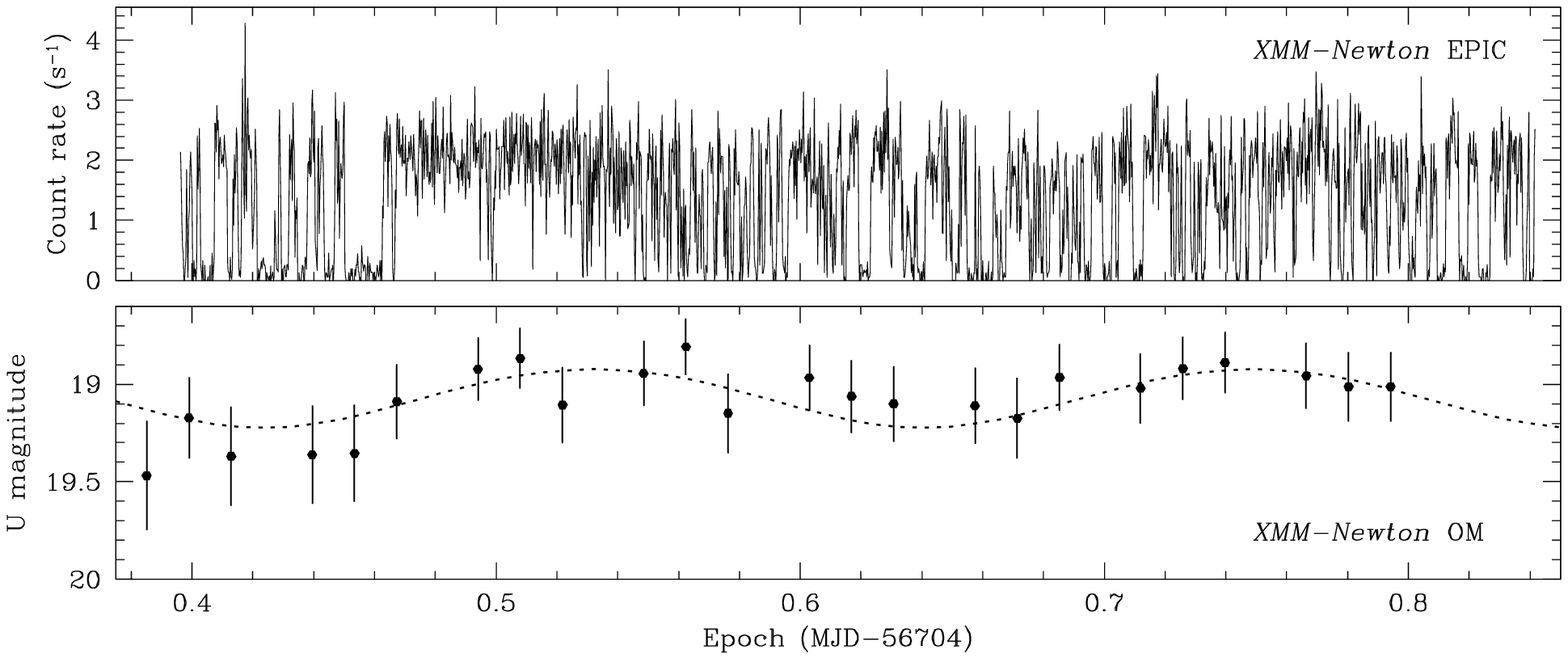}
\includegraphics[height=5.8cm]{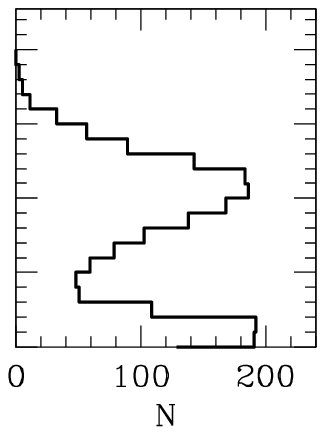}
\end{center}
\caption{Top left: Exposure-corrected, background-subtracted 0.3--10
  keV X-ray light curve of 1RXS J154439.4$-$112820 from the
  \textit{XMM-Newton} EPIC observation obtained by co-adding the
  MOS1/2 and pn light curves binned at a resolution of 20
  seconds. Note the sudden jumps in count rate between $\approx$0.2
  and $\approx$2 counts s$^{-1}$. Top right: Histogram of
    number of 20 second time bins ($N$) as a function of X-ray count
    rate, showing strong bimodality in flux.  Bottom:
  \textit{XMM-Newton} OM $U$ filter photometric light curve binned at
  20 minute resolution. The dotted line shows the best-fit sinusoid
  with period 5.2 hours.}
\end{figure*}

\subsection{MDM Observatory}
We obtained time-series optical photometry of \rxs\ using the MDM
Observatory's 2.4-m Hiltner telescope and 1.3-m McGraw-Hill telescope
on Kitt Peak during the 2014 and 2015 observing seasons.
Table~\ref{tab:log} is a summary of the data obtained on six nights.
The detector was the thinned, backside illuminated SITe CCD
``Templeton''.  It has $1024\times1024$ pixels, with a scale of
$0.\!^{\prime\prime}275$ pixel$^{-1}$ on the 2.4-m and
$0.\!^{\prime\prime}509$ pixel$^{-1}$ on the 1.3-m.  The filter used
was a GG420, to maximize throughput.  In order to reduce the readout
time, the detector was windowed down and the pixels were binned
$2\times2$.  Exposure times were 10~s with a read/prep time of
5\,s, or 30~s with a read/prep time of 3\,s.
Figure~\ref{fig:finder} is a finding chart from the MDM GG420 filter
images, showing the object at position \pos\ from the USNO B1.0
catalog.

\section{X-ray Analysis}
To maximize the photon statistics of the data for the variability
analysis, we extracted a total exposure-corrected and
background-subtracted X-ray light curve of 1RXS J154439.4$-$112820 by
combining the individual MOS1/2 and pn time series over the period
when the three instruments are collecting data simultaneously. This
was done using the {\tt epiclccorr} task in SAS. For this relatively
bright source, removal of periods contaminated by background flares
was not neccessary. We verified that the observed variability in the
background light curve is not present in the background-subtracted
source light curves.

The total 0.3--10 keV light curve, binned at a resolution of
20 s to look for rapid variability, is shown in Figure 2.  The X-ray
light curve exhibits obvious short-timescale, large-amplitude
variability. The X-ray flux alternates irregularly between two
clearly distict modes that differ in flux by a factor of
$\approx$10. Indeed, we find clear bimodality in the X-ray count rate
with two peaks at $\approx$0.2 and $\approx$2 counts s$^{-1}$.  The
transition between the two flux levels is very rapid, with a typical
duration of order 10 s. There is no evidence for periodicity in the
observed variability pattern.

Archival \textit{Swift} XRT data obtained in 2006, 2007, and 2012 also
show X-ray variability comparable to that observed with
\textit{XMM-Newton}. However, due to the brief \textit{Swift}
exposures it is not possible to determine whether the variability
pattern is the same.

We carried out a spectroscopic analysis of the \textit{XMM-Newton}
EPIC data in XSPEC\footnote{Available at
  \url{http://heasarc.nasa.gov/docs/xanadu/xspec/index.html}.}  12.7.1
\citep{Arnaud96}. Due to known issues with the spectral calibration of
the pn instrument in timing mode, we only considered the MOS1/2
data. The time-averaged spectrum of 1RXS J154439.4$-$112820 is
well-described by a simple absorbed power-law. The best fit parameters
are $\Gamma=1.68\pm0.04$ for the spectral photon index, $N_{\rm H}=(1.4
\pm 0.1)\times 10^{21}$ cm$^{-2}$ for the hydrogen column density
along the line of sight, and $\chi^2_{\nu}=0.95$ for 399 degrees of
freedom. The unabsorbed flux in the 0.3--10 keV band is $F_X=(3.51 \pm
0.07) \times 10^{-12}$ erg cm$^{-2}$ s$^{-1}$. All uncertainties
quoted are at a 90\% confidence level.  The derived $N_{\rm H}$ is in
full agreement with the value measured through the Galaxy of $1.3
\times 10^{21}$ cm$^{-2}$ in the direction of this source
\citep{Kalb05}.
Fitting the high and low emission separately (by making a cut
  at 0.9 counts s$^{-1}$ in the light curve from Figure 2) produces
  $\Gamma=1.67\pm0.04$ and $\Gamma=1.97\pm0.28$, implying only
  marginally significant evidence for spectral variations as a
  function of source brightness. As the distance to 1RXS
J154439.4$-$112820 is unconstrained, it is not possible to determine
the X-ray luminosity levels of the high and low modes.

Although the 30 $\mu$s time resolution of the pn permits the search for
X-ray pulsations at millisecond periods, the lack of an orbital period
and ephemeris, combined with the relatively low source count rate,
render a blind periodicity search infeasible at this time.

\section{UV and Optical Variability}
The bottom panel of Figure 2 shows the \textit{XMM-Newton} OM $U$
filter photometric data at a 20 minute resolution, which shows
variability by up to $\approx$0.5 mag.  A fit with a sinusoid
  indicates a periodicity of 5.2 hours. However, it should be noted
  that the low points in the U filter data may not correspond to a
  binary modulation as they could be related to the intervals
  dominated by low mode X-ray emission that occur at the same time.
Due to the faint nature of 1RXS J154439.4$-$112820, it is not possible
to probe variability on timescales of tens of seconds with the OM
data.

\begin{figure}
\begin{center}
\includegraphics[width=0.45\textwidth]{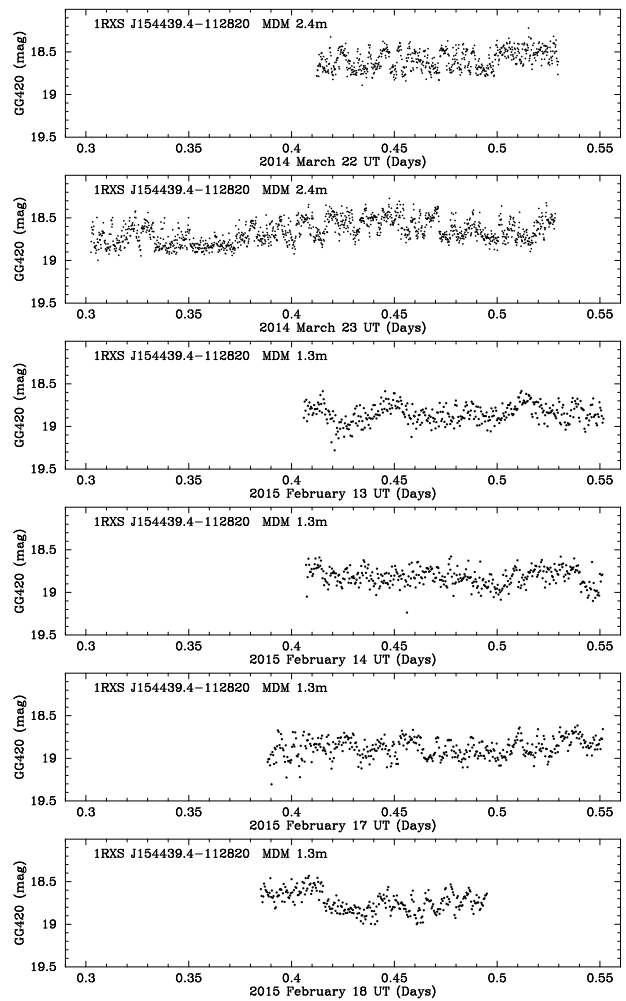}
\end{center}
\caption{MDM light curves of \rxs.
  Magnitude is based on differential photometry
  in the GG420 filter with respect to the comparison star indicated
  in Figure~\ref{fig:finder}.
}
\label{fig:phot}
\end{figure} 

\begin{figure}
\begin{center}
\includegraphics[width=0.45\textwidth]{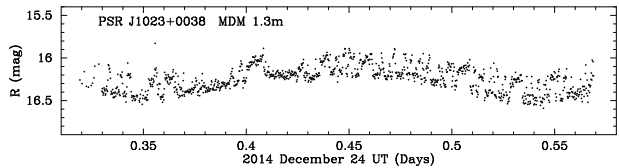}
\end{center}
\caption{MDM 1.3m $R$-band light curve of \psr\ with 13~s cadence.
  The orbital period of the binary is 0.198~d, which is evident
  as the long wave in the light curve. The magnitude
  is based on differential photometry with respect to a
  calibrated comparison star as described in \citet{Bog14b}.
}
\label{fig:psrj1023}
\end{figure}

Differential photometry on the much more sensitive MDM time series was
performed with respect to field stars, and the results are shown in
Figure~\ref{fig:phot}.  Although not calibrated in the GG420 filter,
we adopt the $R$ magnitudes from stars in the USNO B1.0 catalog as an
approximate reference, in particular, $R=14.25$ for the bright star
marked ``C'' on the finding chart.  On this scale, the magnitude of
\rxs\ varies from about 18.3 to 19.1, similar to $R=18.53$ as listed
in the USNO B1.0 catalog.

The nature of the optical variability of \rxs\ observed at MDM is
subtly different from that of most cataclysmic binaries.  Rather than
a random walk or flickering, the brightness appears to be confined
between minimum and maximum values separated by $\approx$$0.5$
magnitudes. This is especially obvious in the higher-quality 2.4m
data. In this regard, it resembles the limit-cycle behavior
occasionally seen from \psr, an example of which is shown in
Figure~\ref{fig:psrj1023}.  There is no obvious candidate for an
orbital or any other period in these light curves of \rxs. A
  periodicity search of the longest light curve on 2014 March 23 shows
  a peak in the power spectrum at $\sim$5.4 hours, but this is unreliable
  because it is comparable to the length of the observation, and it
  cannot be confirmed by any of the shorter time series. In contrast,
in all light curves of \psr, e.g., Figure~\ref{fig:psrj1023}, its
4.75~hr orbital period is clearly manifest as a modulation due to
heating of the side of the companion facing the neutron star.

The Catalina Real-time Transient Survey \citep{Dra09} contains 324
photometric measurements of \rxs\ over a span of 7 years.  We searched
these for periodicity in the range 1--24 hours, such as may arise from
orbital modulation, but did not find a significant detection.

\section{DISCUSSION AND CONCLUSIONS} 
We have presented an X-ray, UV, and optical investigation of 1RXS
J154439.4$-$112820, the most likely counterpart of the \textit{Fermi}
LAT source 3FGL J1544.6$-$1125.  When combined with previous studies,
we can establish the following observational properties of the system:
i) strong emission lines of H and He, suggestive of an accretion disk
\citep[see][]{Mas13}; ii) fast optical variaiblity within limits; iii)
power-law X-ray emission with $\Gamma \approx 1.7$ that exhibits rapid
moding between two distinct X-ray flux levels; iv) persistent
$\gamma$-ray emission with no evidence for a spectral cut-off
\citep[see][]{Fermi15}.  These characteristics are strikingly similar
to what is observed for the transitional MSPs PSR J1824$-$2452I
\citep{Pap13,Pal13,Lin14,Fer14}, PSR J1023+0038
\citep{Stap14,Cot14,Tak14,Bog14b} and XSS J12270--4859
\citep{deM10,deM13,Pap14,Bogd14a,Roy15,Pap15} in their low-luminosity accreting states.
Therefore, 1RXS J154439.4$-$11282 is the counterpart of 3FGL
J1544.6$-$1125 and is almost certainly a binary that hosts an MSP and
is presently in a low-luminosity LMXB state.

As argued by \citet{Del14} for the case of PSR J1023+0038, the
$\gamma$-ray emission seen by \textit{Fermi} LAT is likely generated
in a jet/wind outflow, perhaps driven by propeller-mode accretion.
The peculiar X-ray moding appears to be caused by rapid transitions
between channelled accretion onto the neutron star surface (the high
mode) and no accretion (the low mode).  As with PSR J1023+0038
\citep{Arch14} and XSS J12270--4859 \citep{Pap14}, we predict that
1RXS J154439.4$-$112820 should exhibit coherent X-ray pulsations at
the neutron star spin period during the high flux mode. However, due
to the lack of a precise orbital ephemeris, at present, a blind
periodicity search of the X-ray data is not practical.  In the absence
of evidence for the orbit of \rxs\ in either X-ray or optical light
curves, a spectroscopic period should be obtained from its optical
emission lines.

This system can serve as an additional laboratory for studying
the peculiar phenomenology of transitional MSP systems, especially the
highly unusual X-ray flux switching, and the relation, if any, with
the optical moding we see in 1RXS J154439.4$-$112820. In turn, this
may shed light on the little-understood low-luminosity (``quiescent'')
regime in neutron star X-ray binaries.  In addition, extending the
presently small sample of transitional MSPs has important implication
for understanding key aspects of the evolution of compact neutron star
binaries.

Based on the recent history of the three confirmed LMXB/RMSP
transition objects, we expect 1RXS J154439.4$-$112820 to transform
into a disk-free state in which the rotation-powered millisecond
pulsar activates. Therefore, this object warrants intensive
multi-wavelength monitoring in order to catch such an occurence.

\acknowledgements
A portion of the results presented were based on observations obtained
with \textit{XMM-Newton}, an ESA science mission with instruments and
contributions directly funded by ESA Member States and NASA.  This
work used observations obtained at the MDM Observatory, operated by
Dartmouth College, Columbia University, Ohio State University, Ohio
University, and the University of Michigan. We have made
extensive use of the NASA Astrophysics Data System (ADS) and the arXiv.

Facilities: \textit{XMM}

\end{document}